\UseRawInputEncoding
\documentclass[a4paper,10pt,oneside]{article}
\usepackage[utf8]{inputenc}
\usepackage{icad2023,amsmath,epsfig,times,url,hyperref}
\usepackage[T1]{fontenc}

\title{MIXING LEVELS --- A ROCK MUSIC SPIRIT LEVEL APP}

\name{Tim Ziemer} 
\address{Institute of Systematic Musicology \\
University of Hamburg\\
Hamburg, Germany\\
{\tt tim.ziemer@uni-hamburg.de}} 
\begin{document}
\ninept
\maketitle
\begin{sloppy}
\begin{abstract}
To date, sonification apps are rare. Music apps on the other hand are widely used. Smartphone users like to play with music. In this manuscript, we present \emph{Mixing Levels}, a spirit level sonification based on music mixing. Tilting the smartphone adjusts the volumes of $5$ musical instruments in a rock music loop. Only when perfectly leveled, all instruments in the mix are well-audible. The app is supposed to be useful and fun. Since the app appears like a music mixing console, people have fun to interact with Mixing Levels, so that learning the sonification is a playful experience. 
\end{abstract}

\section{Link to APK File}
\label{sec:apk}
The compiled Mixing Levels app can be found on my 
\href{https://github.com/TimZiemer/mixinglevels-sonictilt-/raw/master/build/app/outputs/flutter-apk/app.apk}{GitHub account}.

\section{Introduction}
\label{sec:intro}
As a systematic musicologist \cite{handbooksm} I carried out research in the field of sound field synthesis for music \cite{book}, music mixing \cite{claudia}, and music information retrieval \cite{fire}, while I worked at the Institute of Systematic Musicology at the University of Hamburg. Then, my academic career brought me to the Bremen Spatial Cognition Center at the University of Bremen, where my focus was sonification for navigation \cite{ijis} and other, mostly medical purposes \cite{schwarz}. A focus on sonification was certainly a good fit, as I could bring in my expertise in sound design, audio programming and digital signal processing, psychoacoustics and auditory cognition in the development and evaluation of audio technology, as well as empiric research. At the same time I learned a lot about cognitive science, medical research and machine learning. But I developed sonifications for dedicated purposes, in which the focus was to achieve the highest possible density of perfectly interpretable information. Musical elements can add a lot concerning acceptance, enjoyment and motivation of sonification applications, but typically at the cost of information density or interpretability. So it was off the table for the purposes of my research projects at the Bremen Spatial Cognition Center. However, being a musicologist by heart, I've always wanted to include music in my sonification research and development. Mixing Levels is my first step in this direction. 

Mixing Levels is a spirit level sonification based on the Sonic Tilt source code, the open source version of Tiltification \cite{tilt}. A major criticism about Tiltification was that some users were not able to interpret the sound that they had never heard before \cite{recommendations}. And they were not eager to learn it, because they disliked the sound. The aim of Mixing Levels is to motivate these users to learn to use a spirit level sonification. When you look at the app market, you can easily see that there are a number of music making and mixing apps around, like \href{https://play.google.com/store/apps/details?id=com.rockrelay.synth.bassline&hl=en&gl=us}{ROCKRELAY: Synthesizer TB 303 Bassline}, \href{https://play.google.com/store/apps/details?id=com.mattfeury.saucillator.android&hl=en&gl=us}{soundandfeury: Saucillator}, \href{https://play.google.com/store/apps/details?id=com.hbi.typeone&hl=en&gl=us}{HOME BAKE INSTRUMENTS: SYNTHESIZER TYPE ONE}, \href{https://play.google.com/store/apps/details?id=com.singlecellsoftware.caustic&hl=en&gl=us}{Single Cell Software: Caustic 3}, \href{https://play.google.com/store/apps/details?id=cx.mccormick.pddroidparty&hl=en&gl=us}{McCormick IT: PdDroidParty} and \href{https://play.google.com/store/apps/details?id=com.csounds.Csound6&hl=en&gl=us}{Irreducible Productions: Csound for Android}. This underlines that the market for music making apps is much larger than the sonification apps market. In contrast to sonification apps, music making apps have already been recognized as a suitable means for education over a decade ago \cite{education}. Many studies concluded that a musical sonification can be informative, motivating, engaging, easy to learn, and raise situation awareness \cite{musicengagement,musicinfo,surgicalsoundtrack}.

In Mixing Levels, while users play around with the music, they learn to interpret the sound. The sound is much more engaging and fun, music is a familiar listening experience, and a reduced information density is fine for a spirit level smartphone app. To be usable without headphones, the sonification should stick to mono and work without the need for frequencies outside the tweeter range between $200$ Hz and $3$ kHz.

\section{Mixing Levels}
\label{sec:format}
The app name \emph{Mixing Levels} is a wordplay: You rotate your smartphone to adjust the mixing levels (i.e., the volumes) of musical instruments, like a music mixing engineer. At the same time, the music mix is supposed to help you level your furniture.

Music recording and mixing has a lot to do with balancing of spectral, spatial, dynamic and temporal aspects \cite{guide,anecdotes,width,1001mixing,1001recording}. In Mixing Levels, you can play music recording/mixing engineer. You need to level your phone in order to balance a musical mix. The music is a loop of $16$ beats duration from my song ``Stay strong'' that I produced in 2008.

The concept of the app is illustrated in Fig. \ref{pic:concept}. You can consider the situation from the viewpoint of a recording engineer: Imagine your smartphone display is a directive microphone, surrounded by musical instruments. On your left, you've got a keyboard, on your right, you've got a piano. Both play the same melody. Between you and the smartphone, you've got an electric guitar, behind the smartphone, you've got a drum kit. Above the smartphone, you've got a squeeky synthesizer, represented by flames. Tilting your smartphone in any direction alters the gain, i.e., the volumes with which all instruments are captured.

When you tilt your smartphone to the right, the piano is focused and will be very loud. When you tilt it towards the left, the volume of the piano will decrease gradually, while the volume of the keyboard increases. To balance the volumes of the two, the smartphone needs to be leveled; neither tilted to the left nor tilted to the right.

When you tilt your smartphone away from you, the drums will be focused and sound very loud. When you gradually tilt it towards you, the volume of the drums will decrease gradually, while the volume of the electric guitar increases. To balance the volumes of the two, the smartphone needs to be leveled; neither tilted towards you nor away from you.

Only when the smartphone is leveled along both dimensions, you can hear all four musical instruments equally loud, and well as the blazing, squeeky synthesizer.

You can also consider the task from the viewpoint of a mixing engineer. The tilt angles control nothing but the gain factors of $5$ audio tracks, according to the rule illustrated in Fig. \ref{pic:gains}: The gain factor for the piano is represented as a green envelope. When the phone is tilted completely to the left, the piano is muted. When tilting it gradually to the right, the gain factor increases linearly. Near $0^\circ$, it reaches a plateau. When tilting it even further to the right, the gain factor rises even further, so that the piano dominates the mix. The gain factor of the keyboard is represented in pink. When tilted completely to the left, the gain factor is at max and the keyboard dominates the mix. While tilting it gradually to the right, the gain factor decreases, until it reaches a plateau near $0^\circ$. When tilted even further to the right, the gain factor decreases further, until the keyboard track is muted. 

The gain factor of the drum kit is represented as a dashed, red envelope. When tilting the smartphone away from you, the gain factor of the drums is too high, so the drums dominate the mix. Tilting it towards you reduces the gain factor. Near $0^\circ$, a plateau is reached, and the drums have an adequate volume. Tilting the display even further towards you, the gain factor reduces further, until the drums become inaudible. The gain factor of the electric guitar behaves in the same manner, but in the opposite direction. When the smartphone is leveled towards you, the gain factor of the electric guitar is at max and the guitar will overly dominate the whole mix. While tilting it away from you, the gain factor will decrease gradually, reaching an adequate plateau near $0^\circ$. When wilting the phone even further away from you, the gain factor will reduce until the electric guitar becomes inaudible.

The volume of the squeeky synthesizer is not controlled continuously by the tilt angle of the phone. Instead, you can imagine the squeeky synthesizer as having a gate effect: Only when a certain threshold is surpassed, the synthesizer is blazing with its full volume. Otherwise, the synthesizer is switched off. The threshold is a tilt angle of $1^\circ$ along both dimensions, represented as dashed, yellow square in the figure.

\begin{figure}[ht]
\centerline{\includegraphics[width=83mm]{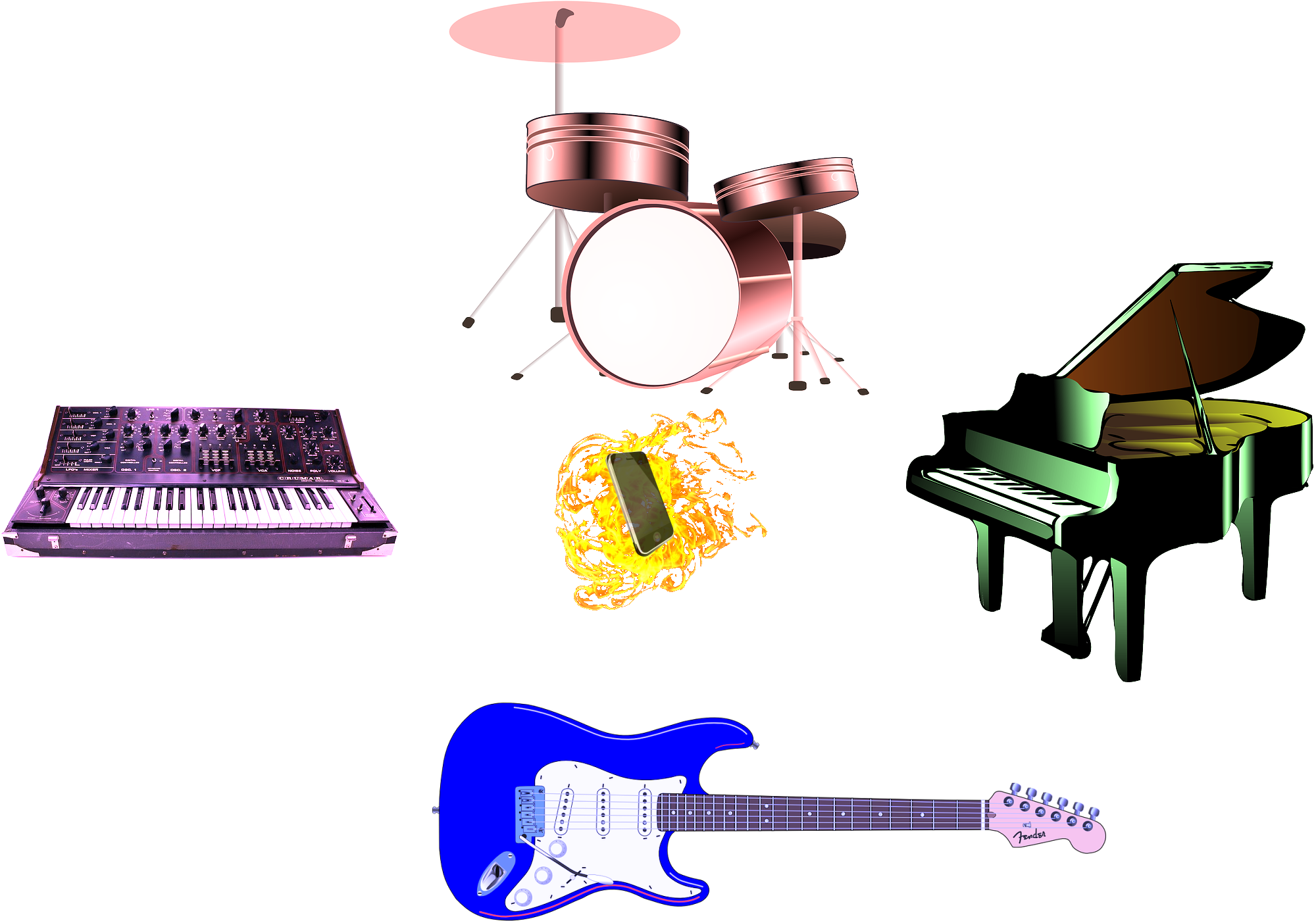}}
	\caption{{\it In Mixing Levels you can figuratively consider your mobile phone display a camera. When tilted to the right, you capture the (green) piano, when tilted to the left, you capture the (pink) keyboard, when tilted away from you, you capture the (red) drums and when tilted towards you, you capture the (blue) electric guitar. Only when perfectly leveled, you capture all instruments equally, and a massive, squeeky (yellow) synthesizer is blazing.}}
	\label{pic:concept}
\end{figure}

\begin{figure}[ht]
\centerline{\includegraphics[width=83mm]{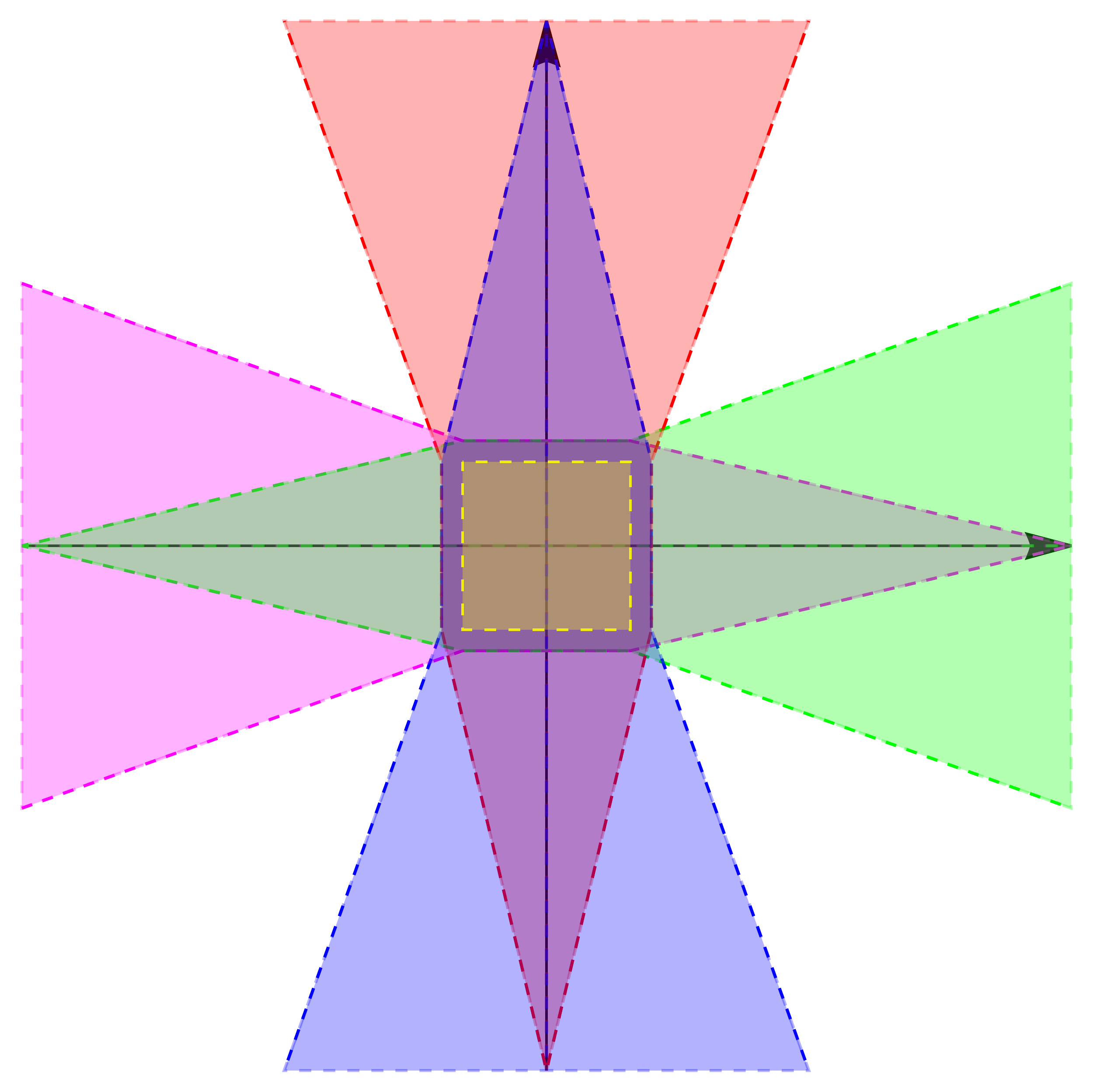}}
	\caption{{\it In Mixing Levels, the tilt angles control the gains of musical instruments. The green envelope represents the gain factor for the piano, the red one for the keyboard, the blue one for the electric guitar and the pink one for the drums. Only when (almost) perfectly leveled, an additional, squeeky synthesizer (yellow) is added.
 }}
	\label{pic:gains}
\end{figure}

A screen capture of Mixing Levels can be found on my YouTube channel under \url{https://youtu.be/bs0xMiljAbs}. Be aware 

\section{Discussion}
\label{discussion}
Mixing Levels uses quite exciting, arousing music. This could be positively shocking for young users, who are used to an overflow of visual and auditory stimuli, e.g., from video games. Users will get the feeling that they are being pushed in the deep end and have to learn how to orientate and make the music work for them instead of being a distractor.

However, many people will want some time to take a breath and explore the sound slowly, step-by-step. They cannot concentrate with such obtrusive music. In everyday-life, people turn down the radio when they need to concentrate \cite{cogn}. And this makes sense, because in most occasions the sound is not the source of information, but a potential distractor. In order to reach the masses, calm music with an ambient character would be preferable. This may also be the more appropriate choice for the task of leveling furniture. A future version may start with a calm piece of music and contain a preset function, in which people can choose their preferred style of music. This would also be a nice feature in terms of UX-design, because a) people can customize the app according to their taste and b) people get even more music to explore and play around with.

\bibliographystyle{IEEEtran}
\bibliography{refs2023.bib}
%
%
%
%

\end{sloppy}
\end{document}